\documentclass[10pt]{article}
\usepackage{amsmath,amssymb}

\numberwithin{equation}{section}

\begin{document}

\title{Existence of relativistic stars in $f(T)$ gravity}
\author{
C. G. B\"ohmer\footnote{c.boehmer@ucl.ac.uk}~,
A. Mussa\footnote{atifahm@math.ucl.ac.uk}~ and
N. Tamanini\footnote{n.tamanini@science.unitn.it}\\
Department of Mathematics and Institute of Origins\\
University College London\\ Gower Street, London, WC1E 6BT, UK
}
\date{\today}
\maketitle

\begin{abstract}
We examine the existence of relativistic stars in $f(T)$ modified gravity and explicitly construct several classes of static perfect fluid solutions. We derive the conservation equation from the complete $f(T)$ gravity field equations and present the differences with its teleparallel counterpart. Firstly, we choose the tetrad field in the diagonal gauge and study the resulting field equations. Some exact solutions are explicitly constructed and it is noted that these solutions have to give a constant torsion scalar. Next, we choose a non diagonal tetrad field which results in field equations similar to those of general relativity. For specific models we are able to construct exact solutions of these field equations. Among those new classes of solutions, we find negative pressure solutions, and an interesting class of polynomial solutions.
\end{abstract}

\section{Introduction}

Modified theories of gravity have become very popular due to their ability to provide an alternative framework to understand dark energy. This is done by modifying the gravitational Lagrangian to become an arbitrary function of its original argument, for instance $f(R)$ instead of $R$ in the Einstein-Hilbert action, see~\cite{Durrer:2008in,DeFelice:2010aj}. In General Relativity and its modifications one uses the metric $g_{\mu\nu}$ and quantities derived from it to describe the gravitational field.

There exists an equivalent formulation of General Relativity based on the idea of parallelism. Initially, Einstein aimed to unify electromagnetism and gravity based on the notion of absolute parallelism~\cite{Unzicker:2005in}, this however failed. Much later, the theory received more attention as an alternative theory of gravity which we now call the teleparallel equivalent of general relativity (TEGR), see~\cite{tegr}.

The idea behind this approach is to consider a more general manifold which contains in addition to curvature a quantity called torsion. The complete Riemann curvature tensor (part without torsion plus a contribution from torsion) is assumed to be zero and therefore one can in principle use either the torsion-free part or the torsion part to describe the gravitational field. The most convenient approach is to work with tetrad fields $e^i{}_{\mu}$ and a so-called  Weitzenb\"{o}ck space, see~\cite{wbock}. The tetrad fields represent fields of orthonormal bases which belong to the tangent space of the manifold. This tangent space is Minkowski space equipped with the metric $\eta_{ij}$ and can be defined at any given point on the manifold. Note that $e^i{}_{\mu}$ has 16 components while the metric has only 10. However, the tetrads are invariant under local Lorentz rotations.

Recently, modifications of TEGR have been studied mainly in the context of cosmology~\cite{Ferraro:2006jd}. This theory is now known as $f(T)$ gravity and is constructed with a generalised Lagrangian~\cite{Bengochea:2008gz}. When compared with $f(R)$ gravity, this modification is particularly appealing since its field equations are of second order and not of fourth order. Note that $f(R)$ modified gravity can also be viewed as a second order system of equations when using the Palatini approach, see again~\cite{Durrer:2008in,DeFelice:2010aj}. Most research on $f(T)$ gravity is devoted to the theory's ability (or inability) to describe the observed accelerated expansion of the universe~\cite{exp}, or comparing theory with observational data~\cite{ob}. In addition, it has been shown that certain $f(T)$ gravity models can provide us with a unification of early time inflation and late time accelerated expansion~\cite{unif}. Models which allow the equation of state to cross the phantom divide have also been found~\cite{pld}. Other lines of research have also been followed, see for instance~\cite{ftothers}.

In two recent papers, static and spherically symmetric solutions were considered~\cite{Deliduman:2011ga,Wang:2011xf} in the context of $f(T)$ gravity. In~\cite{Deliduman:2011ga} it was claimed that relativistic stars in $f(T)$ do not exist, based on the general relativistic conservation equation. In this paper we derive the conservation equation from first principles and show that it agrees with equation~(18) of~\cite{Deliduman:2011ga}. However, this equation should not be compared with its general relativistic analogue. We find solutions with constant torsion scalar $T'=0$, similar to those found in~\cite{Wang:2011xf}. We proceed to consider various simple forms of $g_{\mu\nu}$ to solve the complete set of field equations, thereby showing the existence of relativistic stars. 

\section{Teleparallel gravity and its modifications}

\subsection{Basic equations and action}

The basic variables in the teleparallel approach to general relativity are the tetrad fields $e^i{}_\mu$ where the Greek indices (holonomic) denote the coordinates of the manifold while the Latin indices (anholonomic) denotes the frame. By staggering the frame and the coordinate index, we can use the same symbol for the matrix $e^i{}_{\mu}$ and its inverse. We define
\begin{align}
  e^i{}_{\mu} e_i{}^\nu = \delta^\nu_\mu \,, \qquad
  e^i{}_{\mu} e_j{}^\mu = \delta^i_j \,.
\end{align}
We can define the metric via the tetrads by
\begin{align}
  g_{\mu\nu} = \eta_{ij} e^i{}_{\mu} e^j{}_{\nu} \,,
\end{align}
where $\eta_{ij}=\text{diag}(1,-1,-1,-1)$ is the tangent space metric, which is Minkowski space. Note that the determinant of the metric $g$ is related to the determinant of the tetrad $\sqrt{-g}=\det(e^i{}_{\mu})=e$. The metric $g$ is used to raise and lower coordinate indices and $\eta$ raises and lowers frame indices.

By assuming that the manifold is globally flat, the tetrad fields give rise to a connection defined by
\begin{align}
  \Gamma^{\sigma}_{\mu\nu} = e_i{}^{\sigma} \partial_\nu e^i{}_{\mu}
  = -e^i{}_{\mu} \partial_\nu e_i{}^{\sigma}\,,
  \label{Wbock}
\end{align}
which is the so-called Weitzenb\"{o}ck connection. Note that this connection is not the Levi-Civita connection since it is defined so that its torsion is zero. Since our manifold is flat, the notion of parallelism holds globally and therefore one speaks of absolute parallelism which is a synonym of teleparallelism. 

We define torsion and contortion by
\begin{align}
  T^{\sigma}{}_{\mu\nu} &= \Gamma^{\sigma}{}_{\mu\nu} - \Gamma^{\sigma}{}_{\nu\mu} =
  e_i{}^{\sigma} (\partial_\mu e^i{}_{\nu}-\partial_\nu e^i{}_{\mu}) \,, \\
  K^{\mu\nu}{}_{\sigma} &= -\frac{1}{2}
  (T^{\mu\nu}{}_{\sigma}-T^{\nu\mu}{}_{\sigma}-T_{\sigma}{}^{\mu\nu})\,.
\end{align}              
The contortion tensor can also be defined in terms of the Weitzenb\"{o}ck and Levi-Civita connections. It turns out to be useful to define the tensor $S_{\sigma}{}^{\mu\nu}$ in the following way 
\begin{align}
  S_{\sigma}{}^{\mu\nu} = \frac{1}{2}(K^{\mu\nu}{}_{\sigma} + 
  \delta^\mu_\sigma T^{\rho \nu}{}_{\rho} - \delta_\sigma^\nu T^{\rho\mu}{}_{\rho})\,.
\end{align}
Now we can define a torsion scalar $T$ which is given by 
\begin{align}
  T = S_{\sigma}{}^{\mu\nu} T^{\sigma}{}_{\mu\nu}\,, 
  \label{eqn:torsions}
\end{align}
whose importance becomes clear in a moment.

Due to the flatness of the manifold, we can express the Ricci scalar in the Einstein-Hilbert action in terms of the Weitzenb\"{o}ck connection, or equivalently torsion. This particular combination of torsion terms which appears in this context is the above mentioned $T$. It is thus rather natural to consider modifications of this action based on $f(T)$ where $f$ is an arbitrary function. Let us therefore consider the modified action (with geometrized units $c=G=1$)
\begin{align}
  S = S_{\rm gravity} + S_{\rm matter} = 
  \frac{1}{16\pi} \int e\, f(T) \, d^4x + 
  \int e\, L_{\rm matter} \, d^4x \,.
  \label{eqn:action}
\end{align}

\subsection{Field equations and conservation equation}

Variations of the action~(\ref{eqn:action}) with respect to the tetrads $e^i{}_{\mu}$ gives the field equations of $f(T)$ modified gravity
\begin{align}
  S_i{}^{\mu\nu} f_{TT} \partial_\mu T + e^{-1} \partial_{\mu}(e S_i{}^{\mu\nu}) f_T 
  - T^{\sigma}{}_{\mu i} S_{\sigma}{}^{\nu\mu} f_T
  + \frac{1}{4}e_i{}^{\nu}f = 4\pi \mathcal{T}_{i}{}^{\nu}\,, 
  \label{eqn:field}
\end{align}
where $S_i{}^{\mu\nu}=e_i{}^{\sigma} S_\sigma{}^{\mu\nu}$, $f_T$ and $f_{TT}$ denote the first and second derivatives of $f$ with respect to $T$, see again~\cite{Bengochea:2008gz}. $\mathcal{T}_{\mu\nu}$ is the energy momentum tensor. In what follows we assume $\mathcal{T}_{\mu\nu}$ to be an isotropic perfect fluid which is given by
\begin{align}
  \mathcal{T}_{\mu\nu}= p g_{\mu\nu}-(\rho+p)u_\mu u_\nu\,,
  \label{emtensor}
\end{align}
Conservation of the energy momentum tensor is ensured by the field equations~(\ref{eqn:field}) which we show explicitly in the following. Firstly, we rewrite the field equation in the form
\begin{multline}
  e\delta^\sigma_\rho S_\sigma{}^{\mu\nu}\partial_\mu(T)f_{TT}+e^i{}_{\rho}\partial_\mu(e e_i{}^{\sigma}S_\sigma{}^{\mu\nu})f_T \\-e\delta^\sigma_\rho T^\gamma{}_{\mu\sigma}S_\gamma{}^{\nu\mu}f_T+\frac{e}{4}\delta_\rho^\nu f = 4\pi e\delta^\sigma_\rho \mathcal{T}_\sigma{}^{\nu}\,.
\end{multline}

We introduce the quantity $j_i{}^{\nu}$ to which we refer to as a possible gauge current which represents the energy momentum of the gravitational field
\begin{align}
  j_i{}^{\nu}=-\frac{1}{4\pi}\left(e_i{}^{\sigma}S_\sigma{}^{\mu\nu}\partial_\mu(T) f_{TT} -e_i{}^{\sigma}T^\gamma{}_{\mu\sigma}S_\gamma{}^{\nu\mu}f_T+\frac{1}{4}e_i{}^{\nu}f\right) \,.
  \label{eqn:field1}
\end{align}
If we set $f(T)=T$ then $j_i{}^{\nu}$ reduces to the well known gauge current in teleparallelism~\cite{teleoview}. Using $j_i{}^{\nu}$, equation~(\ref{eqn:field1}) becomes
\begin{align}
  e^i{}_{\rho} \partial_\mu( e e_i{}^{\sigma} S_\sigma{}^{\mu\nu})f_T-4\pi e e^i{}_{\rho}j_i{}^{\nu}=4\pi e\delta^\sigma_\rho \mathcal{T}_\sigma{}^{\nu}\,,
\end{align}
or equivalently 
\begin{align}
 \partial_\mu( ee_i{}^{\sigma}S_\sigma{}^{\mu\nu})f_T-4\pi ej_i{}^{\nu}=4\pi ee_i{}^{\rho}\mathcal{T}_\rho{}^{\nu}\,.
 \label{fieldj}
\end{align}

Secondly, we now take the derivative of~(\ref{fieldj}) with respect to $x^\nu$. The antisymmetry of $S_\sigma{}^{\mu\nu}$ (in the last pair of indices $S_\sigma{}^{\mu\nu}=-S_\sigma{}^{\nu\mu}$) implies the conservation equation
\begin{align}
  4\pi\partial_\nu(e(j_i^{\nu}+\mathcal{T}_i{}^{\nu})) = \partial_\mu(eS_i{}^{\mu\nu})\partial_\nu f_T\,.
  \label{generalcons}
\end{align}
We immediately observe a few interesting facts. The presence of $f(T)$ in the action affects the conservation equation in two different ways. On the one hand, the gauge current changes and on the other hand, also the right hand side is affected. When $f(T)=T$ we recover the well-known conservation equation of TEGR~\cite{teleoview}. In the following section we discuss static and spherically symmetric solutions in $f(T)$ gravity and it will turn out that the contributions from the right-hand side term are crucial.

\section{Solutions with diagonal tetrad}

\subsection{Field equations}

Consider the static spherically symmetric metric 
\begin{align}
  ds^2 = e^{a(r)} dt^2 - e^{b(r)} dr^2 -R(r)^2 d\Omega^2\,,
  \label{metric1}
\end{align}
where $d\Omega^2 = d\theta^2 + \sin^2\negmedspace\theta\, d\varphi^2$ and where $a$, $b$ and $R$ are three unknown functions. One possible tetrad field (we can make arbitrary Lorentz transformations to the tetrads without changing the metric) can be written as
\begin{align}
  e^i{}_{\mu} = \text{diag}(e^{a(r)/2},e^{b(r)/2},R(r),R(r)\sin \theta)\,,
  \label{tetrad1}
\end{align}
to which we refer to as the diagonal gauge. The determinant of this diagonal matrix is the product of its elements so that we find $e=e^{(a+b)/2}R^2\sin\theta$. Based on this tetrad field, we can now write out explicitly the $f(T)$ field equations~(\ref{eqn:field}) in component form. To do so, we firstly compute the torsion scalar~(\ref{eqn:torsions}) and its derivative which become
\begin{align}
  T(r) &= 2e^{-b}\frac{R'}{R} \left(a'+\frac{R'}{R}\right)\,,
  &\nonumber\\
  T'(r) &=-2e^{-b}\left( -a''\frac{R'}{R} + a'\frac{R''}{R} - \frac{2R'R''}{R^2} + \frac{R'^3}{R^3}\right) - T \left( b'+\frac{R'}{R}\right) \,.
  \label{Tscalar}
\end{align}
Inserting this and the components of the tensors $S$ and $T$ in to the diagonal components of the equation~(\ref{eqn:field}) yields
\begin{align}
  4 \pi \rho&=e^{-b}\frac{R'}{R}T' f_{TT}+\left(T-\frac{ 1}{R^2}-e^{-b}\left((a'+b')\frac{R'}{R}-2\frac{R''}{R}\right)\right)\frac{f_T}{2}-\frac{f}{4} \,,
  \label{field:t}\\
  4\pi p&= \left(\frac{1}{R^2}-T\right)\frac{f_T}{2}+\frac{f}{4} \,,
  \label{field:r}\\
  4 \pi p&= -\frac{e^{-b}}{2}\left(\frac{a'}{2}+\frac{R'}{R}\right)T'f_{TT}\nonumber\\
  &-\left( \frac{T}{2}+e^{-b}\left(\frac{R''}{ R}+\frac{a''}{2}+\left(\frac{a'}{4}+\frac{R'}{2 R}\right)(a'-b')\right)\right)\frac{f_T}{2} +\frac{f}{4}\,.
  \label{field:theta}
\end{align}
In General Relativity the off diagonal components of the field equations vanish, but in $f(T)$ gravity a diagonal tetrad field~(\ref{tetrad1}) gives rise to an extra equation\footnote{We are deeply indebted to Franco Fiorini for pointing this out to us and thereby correcting our previously incomplete field equations.}. This is the $(r,\theta)$ component:
\begin{multline*}
  \frac{e^{-\frac{3 b}{2}} \cot (\theta ) f_{TT}}{R^2} 
  \Bigl(-a' \frac{R''}{R} + (a'+b')\frac{R'^2}{R^2} \\ +
  2 \frac{R'^3}{R^3} + \frac{R'}{R} \left(\left(a' b'-a''\right)-2 \frac{R''}{R}\right)\Bigr)=0\,,
\end{multline*}
which can be written in terms of $T'$ and gives
\begin{align}
  \frac{e^{-b/2}\cot\theta}{2R^2}T'f_{TT}=0\,.
  \label{field:extra}
\end{align}
This implies that all solutions satisfy either $f_{TT} = 0$ or $T'=0$, where the former reduces the theory to TEGR.

The above field equations~(\ref{field:t})--(\ref{field:extra}) provide us with four independent equations for six unknown quantities, namely $a(r)$, $b(r)$, $R(r)$, $\rho(r)$, $p(r)$ and $f(T)$. Hence, this system of equations is under-determined, and in order to find solutions, we will need to make some reasonable assumptions to reduce the number of unknown functions to four. 

The most physical approach would be to prescribe an equation of state $\rho=\rho(p)$ which relates the energy density and the pressure, and to prescribe the function $f(T)$. However, even in general relativity it turns out to be difficult to find explicit solutions for a realistic equation of state. Moreover, prescribing a possibly complicated function $f(T)$ will make the field equations even harder to solve as they would contain more nonlinear terms. Thus, it turns out to be best to follow alternative routes. We will either make assumptions about the metric functions, or we will choose useful combinations of terms which simplify the field equations. 

\subsection{Conservation equation II}

Based on the field equations~(\ref{field:t})--(\ref{field:extra}), we will re-derive the conservation directly from them, without any reference to the previous results. In doing so, we will have two independent derivations of the conservation equation in the static and spherically symmetric setting based on a diagonal tetrad. Our approach here follows closely the well known derivation of the conservation equation in general relativity, see for instance~\cite{Tolman:1939jz}. In order to avoid working in TEGR, let us assume $T'=0$ in what follows.

Differentiating~(\ref{field:r}) gives
\begin{align}
  4\pi p'(r) = -\frac{R'}{R^3}f_T\,.
  \label{step1}
\end{align}
An expression for $\rho+p$ can be obtained using~(\ref{field:t}) and~(\ref{field:r})
\begin{align}
  4\pi(\rho+p) = -\frac{e^{-b}}{2R}\left(R'(a'+b')-2R''\right)f_T\,,
  \label{step2a}
\end{align}
while isotropy of the pressure implies
\begin{align}
  \left(\frac{1}{2R^2}-\frac{T}{4}\right)f_T+\frac{e^{-b}}{2}\left(\frac{R''}{ R}+\frac{a''}{2}+\left(\frac{a'}{4}+\frac{R'}{2 R}\right)(a'-b')\right)f_T = 0\,.
  \label{step3a}
\end{align}
Let us multiply~(\ref{step2a}) by $a'/2$ and~(\ref{step3a}) by $2R'/R$ and subtract the two resulting equations. Since $T'=0$ we arrive at 
\begin{align}
  2\pi a'(\rho+p) =\frac{R'}{R^3}f_T\,.
\label{step4}
\end{align}
Substituting~(\ref{step4}) into~(\ref{step1}) leads to the $f(T)$ conservation equation in a static and spherically symmetric spacetime with a diagonal tetrad field
\begin{align}
  4\pi p' + 2\pi a'(\rho+p) = 0 \,.
  \label{ssscons}
\end{align}
Notice that the right-hand side of this equation always vanishes because of the off-diagonal equation~(\ref{field:extra}) which enforces $f_{TT}T'=0$. Thus in $f(T)$ gravity the static and spherically symmetric conservation equation in diagonal gauge coincides with its general relativistic counterpart, meaning that no physical differences can arise from it, compare with~\cite{Deliduman:2011ga}.  

At this point we would like to recall the general conservation equation~(\ref{generalcons}) derived in the previous section
\begin{align*}
  -\partial_\mu(eS_i{}^{\mu\nu})\partial_\nu(f_T)+4\pi \partial_\nu(e \mathcal{T}_i{}^{\nu})+4\pi \partial_\nu( e j_i{}^{\nu}) = 0\,.
\end{align*}
Then, given the tetrad~(\ref{tetrad1}), we find
\begin{align}
  \frac{e^{a/2} \sin\theta T' f_{TT}}{2} - 4\pi e^{a/2} r^2\sin\theta
  \left( p'+\frac{a'p}{2}+\frac{2p}{r}+\frac{a'\rho}{2}-\frac{2p}{r}\right) 
  &=0\,.
\end{align}
By rearranging and inserting $T'=0$ we see that this agrees with the conservation equation~(\ref{ssscons}) derived above. We are now ready to construct perfect fluid solutions for a constant torsion scalar explicitly.

\subsection{Solutions with $T=0$}
\label{sec:T=0diag}

Let us start by analysing the probably simplest solution. Let us assume that $T=0$ and insert this into~(\ref{Tscalar}). This fixes the function $a(r)$ and we obtain
\begin{align}
  a' &= -\frac{R'}{R} \,, \qquad a(r) = \ln(c_1/R(r))\,, 
\end{align}
where $c_1$ is a constant of integration. Note that due to $T=0$, $f$ and its derivatives are now constants. Rewriting the field equations with this gives
\begin{align}
  4\pi\rho &= -\left(\frac{1}{2R^2} - \frac{e^{-b}}{2} \left( \frac{R'}{R}\left(\frac{R'}{R}+b'\right)+2\frac{R''}{R}\right)\right) f_T(0) - \frac{f(0)}{4}\,, 
  \label{t0rho}\\
  4\pi p &= \frac{f_T(0)}{2R^2}+\frac{f(0)}{4}\,, 
  \label{t0p}\\
  4\pi p &= -\frac{e^{-b}}{4}\left(\frac{R''}{R}+\frac{R'^2}{2R^2}-\frac{b'}{2}\frac{R'}{R}\right)f_T(0)+ \frac{f(0)}{4}\,,
  \label{t0p2}
\end{align}
while the conservation~(\ref{ssscons}) simplifies to 
\begin{align}
  4\pi p' + \frac{2\pi R'}{R}(\rho+p) = 0\,.
\end{align}
Isotropy of the pressure implies that $b$ has to satisfy the differential equation
\begin{align}
  \frac{e^{-b}}{2} 
  \left(-\frac{b'}{2} \frac{R'}{R} + \frac{R''}{R} + \frac{R'^2}{2 R^2}\right)
  + \frac{1}{R^2} = 0 \,.
\end{align}
We can solve for $b$
\begin{align}
  b = -\ln \left(\frac{c_2-4R}{R R'^2}\right)\,,
\end{align}
where $c_2$ is another constant of integration. Therefore, the metric coefficients are given by
\begin{align}
  e^{a}=\frac{c_1}{R}\,, \quad e^{b} = \frac{R R'^2}{c_2-4R}\,,
 \label{006}
\end{align}
and we arrive at the metric
\begin{align}
ds^2= \frac{c_1}{R} dt^2 -\frac{R R'^2}{c_2-4R} dr^2 - R^2d\Omega^2\,.
\label{metricT=0}
\end{align}
Having determined both metric functions, we can now attempt to solve the remaining equations. If we set $f_T(T=0)=0$, then the field equations simply imply
\begin{align}
  \rho_0=-p_0=\frac{f_0}{16 \pi}\,,
\end{align}
thus we have found all unknown functions. Note that $f(T)$ is arbitrary in the sense that only its value at the origin (and of its derivative) is of importance. Note that this excludes those $f(T)$ which become singular as $T \rightarrow 0$. This solution yields a constant energy density $\rho_0$ and pressure $p_0$, obeying the dark energy equation of state. Notice that the metric~(\ref{metricT=0}) has a singularity when $R \rightarrow 0$. Moreover, we expect a coordinate singularity at $R=c_2/4$ and it might be interesting to study this surface in more detail.

\subsection{Solutions with $T'=0$}

Rather than assuming $T=0$, we now assume $T'=0$ which is equivalent to assuming $T=\text{constant}$. These solutions are rather complicated but simplify substantially when considering the $R(r)=r$ case. Let 
\begin{align*}
  T(r)=T_0=\text{constant}\,,
\end{align*}
then the conservation equation~(\ref{ssscons}) is again given by
\begin{align}
  p'=-\frac{a'}{2}(\rho+p) \,.
\end{align}
The field equation~(\ref{field:r}) becomes
\begin{align}
  4\pi p=\frac{T_0}{2}f_T(T_0)-\frac{f(T_0)}{4}+\frac{f_T(T_0)}{2r^2} \,,
\end{align}
where $f(T_0)$ and $f_T(T_0)$ are constants. As $r\rightarrow\infty$ we have
\begin{align}
  4\pi p_\infty = \frac{T_0}{2}f_T(T_0)-\frac{f(T_0)}{4} \,,
  \label{diffp}
\end{align}
with $p_\infty$ denoting the value of the pressure at infinity.

Once we fix $p_\infty$,~(\ref{diffp}) can be viewed as a differential equations for $f(T_0)$ whose general solution is
\begin{align}
  f(T_0) = \tilde{f} \sqrt{T_0} - 16\pi p_\infty \,,
\end{align}
where $\tilde{f}$ is a constant of integration. Note that in this model $p_\infty$ plays the role of a cosmological constant. However, though it represents a very simple spherically symmetric solution, it seems not to be of general interest in cosmology due to its incompatibility with standard teleparallel gravity.

Using the equation for $T$,
\begin{align}
  T_0=\frac{2e^{-b}}{r^2} \left(a'r+1\right) \,,
  \label{007}
\end{align}
in order to write $a'$ in terms of $b$ and comparing~(\ref{field:t}) and~(\ref{field:r}) we obtain the following differential equation for $b$,
\begin{align}
  \frac{Y'}{2r}\left(1-\frac{1}{2Y}r^2T_0\right) +\frac{Y}{2r^2}\left(1+\frac{2}{Y^2}r^4T_0^2\right) +\frac{2}{r^2}\left(1+\frac{3}{4}r^2T_0\right) =0 \,,
  \label{005}
\end{align}
where $Y(r)\equiv e^{-b(r)}$. An analytical solution to this differential equation seems hard (if not impossible) to find. However under certain approximations we can arrive to some solutions.

First, consider the $r^2T_0\ll 1$ regime where~(\ref{005}) reduces to
\begin{align}
  \frac{Y'}{2r} +\frac{Y}{2r^2} +\frac{2}{r^2} =0 \,.
\end{align}
The solution is then
\begin{align}
  Y(r)=e^{-b(r)}=\frac{c_1}{r}-4 \,,
\end{align}
which of course coincides with~(\ref{006}), the solution for $T=0$. To find also $a(r)$ we can use~(\ref{007}), which in the regime $r^2T_0\ll 1$ gives again the solution~(\ref{006}). Thus, at zeroth order in $r^2T_0$ the general solution matches the $T=0$ solution, as one would certainly expect.

On the other hand, we can analyze the opposite regime: $r^2T_0\gg 1$. Equation~(\ref{005}) becomes now
\begin{align}
  \frac{r}{4}Y' +\frac{3}{2}Y +r^2T_0 =0 \,,
\end{align}
and the solution is given by
\begin{align}
  Y(r)=e^{b(r)}=\frac{k_1}{r^6}-\frac{T_0}{2}r^2 \,,
\end{align}
where $k_1$ is a constant.
Again, to find $a(r)$ we go back to~(\ref{007}) which for $r^2T_0\gg 1$ reduces to
\begin{align}
  a'\simeq \frac{rT_0}{2e^{-b}} \,,
\end{align}
and gives
\begin{align}
  a(r)=-\frac{1}{8} \ln\left(\frac{r^8T_0-2k_1}{k_2}\right) \,,
\end{align}
with $k_2$ another integration constant. Thus the metric in the regime $r^2T_0\gg 1$ is
\begin{align}
  ds^2 = \left(\frac{k_2}{T_0r^8-2k_1}\right)^{1/8} dt^2 - \frac{2k_1-T_0r^8}{2r^6} \,dr^2 -r^2 d\Omega^2\,.
\end{align}
This metric is singular for $r\rightarrow 0$ and $r\rightarrow \infty$ and presents an horizon for $r=(2k_1/T_0)^{1/8}$.

\subsection{Triviality of the Einstein static universe}

Consider the case $R(r)=r$ and the metric~(\ref{metric1}) with $e^{a(r)}$ and $e^{b(r)}$ fixed such that
\begin{align}
  ds^2= dt^2-\frac{1}{1-kr^2}dr^2-r^2d\Omega ^2\,,
  \label{metric2}
\end{align} 
so that the corresponding diagonal tetrad field is given by
\begin{align}
  e^i{}_{\mu}=\text{diag}\left(1,\frac{1}{\sqrt{1-kr^2}},r,r\sin\theta\right) \,.
\end{align}
Since we have now chosen three functions, the system of equations is closed. For this choice the torsion scalar reads
\begin{align}
  T=-\frac{2(1-kr^2)}{r^2}\,, \quad 
  T'=\frac{4}{r^3} \,,
  \label{EstaticT}
\end{align}
and the field equations~(\ref{field:t})--(\ref{field:theta}) become
\begin{align}
  4\pi\rho_0 &= \frac{4(1-kr^2)}{r^4}f_{TT}+\left(\frac{1}{2r^2}-k\right)f_T+\frac{f}{4}\,,\\
  4\pi p &= -\left(\frac{1}{2r^2}-k \right)f_T-\frac{f}{4}\,,\\
  4\pi p &= -\frac{2(1-kr^2)}{r^4}f_{TT}-\left(\frac{1}{2r^2}-k \right)f_T-\frac{f}{4}\,,\\
  0 &= \frac{e^{-b/2}\cot\theta}{2R^2}T'f_{TT}\,.
\end{align}
The last field equation, and the isotropy of the pressure respectively imply 
\begin{align}
  T'f_{TT}=0\,, \qquad -\frac{2(1-kr^2)}{r^4}f_{TT}=0\,.
  \label{step1ES}
\end{align}
Since $1-kr^2$ cannot be zero, this can only be satisfied if $f_{TT}=0$ which takes us back to TEGR. Note that we cannot achieve $T'=0$ due to~(\ref{EstaticT}). 

\section{Solutions with off diagonal tetrad}

\subsection{Field equations}

Consider again the static spherically symmetric metric 
\begin{align}
  ds^2 = e^{a(r)} dt^2 - e^{b(r)} dr^2 -R(r)^2 d\Omega^2\,,
  \label{metric1off}
\end{align}
where $d\Omega^2 = d\theta^2 + \sin^2\negmedspace\theta\, d\varphi^2$ and where $a$, $b$ and $R$ are three unknown functions. Another possible tetrad field can be written as
\begin{align}
  e^i{}_{\mu} = \left(
\begin{array}{cccc}
 e^{a/2} & 0 & 0 & 0 \\
 0 & e^{b/2} \sin\theta\cos\phi  & R\cos \theta\cos\phi & -R\sin \theta\sin \phi \\
 0 & e^{b/2} \sin\theta\sin\phi & R\cos \theta\sin\phi& R\sin\theta \cos\phi \\
 0 & e^{b/2} \cos \theta  & -R \sin\theta & 0
\end{array}
\right)\,,
  \label{tetradoff}
\end{align}
see for instance~\cite{teleoview}. The determinant of $e^i{}_\mu$ is $e=e^{(a+b)/2}R(r)^2\sin\theta$. The torsion scalar and its derivative are
\begin{align}
  T(r) = \frac{2 e^{-b} \left(e^{b/2}-R'\right)\left(e^{b/2}-R'-R a '\right)}{R^2}\,,
\end{align}
\begin{multline}
  T'(r) = -\frac{e^{-b/2}}{R^2}\left(4 R''+2R'\left(a'-b'\right)+R \left(2a''-a'b'\right)\right) \\
  +\frac{2e^{-b}}{R^2}\left(R R''a'+R'^2 \left(a'-b'\right)+R' \left(2 R''+R\left(a''-a'b'\right)\right)\right)-\frac{2R'T}{R}\,.
  \label{Tscalaroff}
\end{multline}
Inserting this and the components of the tensors $S$ and $T$ into the equation~(\ref{eqn:field}) yields
\begin{multline}
  4\pi\rho = \frac{e^{-b/2}}{R}(R'e^{-b/2}-1) T'f_{TT}+\left(\frac{T}{4}-\frac{1}{2R^2}\right)f_T \\
  +\frac{e^{-b}}{2 R^2} \left(2 RR''-R R'b '+R'^2\right)f_T-\frac{f}{4}\,,
  \label{field:toff}
\end{multline}
\begin{align}
  4\pi p =\left( \frac{1}{2R^2}-\frac{T}{4}-\frac{e^{-b}}{2R^2} R'(R'+Ra')\right)f_T+\frac{f}{4}\,,\mbox{\qquad\qquad\qquad\qquad\qquad\mbox{}}
  \label{field:roff}
\end{align}
\begin{multline}
  4\pi p = -\frac{e^{-b}}{2}\left(\frac{a'}{2}+\frac{R'}{R}-\frac{e^{b/2}}{R}\right)T' f_{TT} \\
  -f_T\left( \frac{T}{4}+\frac{e^{-b}}{2 R} \left(R''+\left(\frac{R'}{2}+\frac{Ra'}{4}\right) \left(a'-b'\right)+\frac{Ra''}{2}\right)\right)+\frac{f}{4}\,.
  \label{field:thetaoff}
\end{multline}

The above field equations~(\ref{field:toff})--(\ref{field:thetaoff}) give three independent equations for our six unknown quantities. As before, this system of equations is under-determined, we will reduce the number of unknown functions by assuming suitable conditions. Note that there is no equation enforcing the constancy of the torsion scalar in this non diagonal gauge. 

\subsection{Conservation equation III}

We will derive the conservation equation one more time, now for the off diagonal tetrad. Taking the derivative of~(\ref{field:roff}) gives
\begin{multline}
  4\pi p'(r)= \frac{e^{-b}}{2 R^2}\left(2 R' \left(e^{b/2}-Ra'-R'\right)+Ra' e^{b/2}\right)T'f_{TT}-\frac{R'}{R^3}f_T \\
  -\frac{e^{-b}}{2R}\left(\frac{R'}{R} \left(R'(a'+b')-2 R''\right)+\left(a'\left(b'R'-R''\right)-a'' R'\right)+\frac{2R'^3}{R^2}\right)f_T \,.
\label{step1off}
\end{multline}
Next, we take a combination of the field equations~(\ref{field:toff}) and~(\ref{field:roff}) to obtain
\begin{multline}
  4\pi(\rho+p)=\frac{e^{-b/2}}{R}\left(R'e^{-b/2}-1\right)T'f_{TT}-\frac{R'e^{-b}}{2R^2}\left(R'+Ra'\right)f_T\\
  +\frac{e^{-b}}{2 R^2} \left(2 RR''-R R'b '+R'^2\right)f_T \,.
\label{step2off}
\end{multline}
Isotropy of pressure allows us to write
\begin{multline}
  -\frac{e^{-b}}{2}\left(\frac{a'}{2}+\frac{R'}{R}-\frac{e^{b/2}}{R}\right)T' f_{TT}+\frac{R'e^{-b}}{2R^2}(R'+Ra')f_T\\
  -\frac{e^{-b}}{2 R} \left(R''+\left(\frac{R'}{2}+\frac{Ra'}{4}\right) \left(a'-b'\right)+\frac{Ra''}{2}\right)f_T-\frac{1}{2R^2}f_T=0 \,.
\label{step3off}
\end{multline}
We now multiply~(\ref{step2off}) by $a'/2$ and~(\ref{step3off}) by $2R'/R$ and subtract the two resulting equations
\begin{multline}
  2\pi a'(\rho+p)=-\frac{e^{-b}}{2R^2}\left(2R'\left( e^{b/2}-Ra'-R'\right)+Ra'e^{b/2} \right)T'f_{TT}+\frac{R'}{R^3}f_T\\
  +\frac{e^{-b}}{2R}\left(\frac{R'}{R} \left(R'(a'+b')-2 R''\right)+\left(a'\left(b'R'-R''\right)-a'' R'\right)+\frac{2R'^3}{R^2}\right)f_T\,,
\end{multline}
comparing this with equation~(\ref{step1off}), we obtain the conservation equation for the off diagonal tetrad field~(\ref{tetradoff})
\begin{align}
  p' + \frac{a'}{2}(\rho+p) = 0\,.\label{sssconoff}
\end{align}
It is surprising that one recovers the general relativistic conservation equation directly in this approach. This seems to indicate that the non diagonal tetrad~(\ref{tetradoff}) is a very good starting point to study static and spherically symmetric perfect fluid solutions in $f(T)$ gravity. It also eliminate the problematic field equation in diagonal gauge which posed stringent constraints on the possible solutions.

\subsection{Solutions with $b=0$}

Consider our metric~(\ref{metric1off}) and tetrad field~(\ref{tetradoff}) with $b(r)=0$ and $R(r)=r$, then we have the metric
\begin{align}
  ds^2=e^{a(r)}dt^2-dr^2-r^2d\Omega^2\,,
\end{align}
and for all values of $a$ and $R$ the torsion scalar is given by
\begin{align}
  T(r)=0\,.
\end{align}
Again, since we have a vanishing torsion scalar, $f$ and its derivatives are constant. The field equations simplify to 
\begin{align}
  4\pi \rho&=-\frac{f(0)}{4} \,,\\
  4\pi p &=-\frac{a'}{2r}f_T(0)+\frac{f(0)}{4} \,,\\
  4\pi p&=-\left(\frac{a''}{4}+\frac{a'}{4}\left(\frac{a'}{2}+\frac{1}{r}\right)\right)f_T(0) +\frac{f(0)}{4}\,. 
\end{align}
Isotropy of the pressure yields
\begin{align}
  a'' + \frac{a'^2}{2} - \frac{a'}{r} = 0\,,
\end{align}
which we can solve for $a$ and find
\begin{align}
  a = 2\ln(r^2+c_1)+\ln c_2\,,
\end{align}
where $c_1$ and $c_2$ are constants of integration. Thus we arrive at the metric
\begin{align}
  ds^2= c_2 (r^2 + c_1)^2 dt^2- dr^2- r^2d\Omega^2\,.
  \label{metricb0}
\end{align}
Since $f$ and $f_T$ are constants we can label $f(0)=f_1$ and $f_T(0)=f_2$. Using this and the metric coefficient $a$, we can write the field equations more explicitly
\begin{align}
  4\pi \rho&=-\frac{f_1}{4} \,,\label{fieldb1}\\
  4\pi p &=-\frac{2}{r^2+c_1}f_2+\frac{f_1}{4} \,,\label{fieldb2}\\
  4\pi p&=-\frac{2}{r^2+c_1}f_2 +\frac{f_1}{4}\label{fieldb3}\,.
\end{align}
Immediately, we see that $\rho$ and $p$ are given by
\begin{align}
  \rho_0 &= -\frac{f_1}{16\pi} \,,\\
  p & =-\frac{1}{2\pi(r^2+c_1)}f_2+\frac{f_1}{16\pi} \,.
\end{align}
Notice that the pressure is regular everywhere provided that $c_1 > 0$.

\subsection{Solutions with $f_T(T=0)=0$}
\label{sec:f'(0)=0 solutions}

We now build a class of solutions for $R(r)=r$. Constraining the metric function $b$ to be
\begin{align}
  b=2\ln\left(1+r\,a'\right)\,,
  \label{009}
\end{align}
leads to $T=0$ for all $a$. The field equations~(\ref{field:toff})-(\ref{field:thetaoff}) reduce to
\begin{align}
  4 \pi \rho &=\frac{f_T(0) \left(2 r a''+a' \left(r a' \left(r a'+3\right)+4\right)\right)}{2 r \left(r a'+1\right)^3}-\frac{f(0)}{4} \,,\label{010}\\
  4 \pi p &=\frac{a' f_T(0)}{2 \left(r^2 a'+r\right)}+\frac{f(0)}{4} \,,\label{011}\\
  4\pi p &=-\frac{f_T(0) \left(a' \left(r a'-1\right) \left(r a'+2\right)-2 r a''\right)}{8 r \left(r a'+1\right)^3}+\frac{f(0)}{4} \,.\label{012}
\end{align}
If we further assume $f_T(0)=0$, so that we only consider particular $f(T)$ models, these field equations imply
\begin{align}
  p=-\rho =\frac{f(0)}{16\pi} \,.
\end{align}
The constraint $f_T(0)=0$ is satisfied by a large amount of $f(T)$ models. For example, we can take $f(T)=T^n$ (with $n\neq 1$) or $f(T)=\cos(kT)$. The first gives $p=\rho=0$, while the second gives $p=-\rho=1/(16\pi)$. For all these models we have an infinite class of solutions given by choosing an arbitrary form for $a(r)$ and constraining $b(r)$ with the condition~(\ref{009}). For instance, considering a Schwarzschild form for $a(r)$,
\begin{align}
  a(r)=\ln\left(1-K/r\right)\,,
\end{align}
we find
\begin{align}
  b(r)=-2\ln\left(1-K/r\right)\,,
\end{align}
with $K$ being a constant. The metric becomes
\begin{align}
  ds^2= \left(1-K/r\right)dt^2-\left(1-K/r\right)^{-2}dr^2-r^2d\Omega^2\,.
\end{align}
which is very similar, but not equal, to the Schwarzschild solution.

We can compare these solutions with the ones we found in Sec.~(\ref{sec:T=0diag}) with a diagonal tetrad. We notice that assuming $f_T(0)=0$ implies a Dark Energy equation of state in both cases. This suggests that, for these models, such a characteristic does not depend upon the choice of the tetrad.

\subsection{A negative pressure solution}

Following the route of Sec.~(\ref{sec:f'(0)=0 solutions}) we now build a new interesting solution which applies to every $f(T)$ model. We still assume $R(r)=r$ and $b(r)$ as in~(\ref{009}) which again gives $T=0$. As a particular form for $a(r)$ we consider
\begin{align}
  a(r)=-\frac{4}{5}\ln(k\,r) \,,
\end{align}
where $k$ is a constant. This implies $b(r)=-2\ln 5$, corresponding to the following metric,
\begin{align}
  ds^2= (k\,r)^{-4/5}dt^2-\frac{1}{25}dr^2-r^2d\Omega^2\,.
\end{align}
Note that the factor $1/25$ in front of $dr^2$ is physically unimportant. This is a special solution inasmuch as it renders both the field equations~(\ref{011}) and~(\ref{012}) equal to
\begin{align}
  4\pi p=\frac{1}{4}f(0)-\frac{2}{r^2}f_T(0) \,,
\end{align}
without considering any constraint on $f(T)$, which means that it works for all the $f(T)$ models. The field equation~(\ref{010}) becomes
\begin{align}
  4\pi\rho=-\frac{1}{4}f(0)+\frac{12}{r^2}f_T(0) \,.
\end{align}
Note that if we require $\rho$ to be positive everywhere we must have $f(0)<0$ and $f_T(0)>0$. This implies that $p$ has to be negative for every $r$ meaning that this is in general a negative pressure solution. Solutions of this type might be interesting in the context of dark energy or may have applications for wormhole like solutions. 

\subsection{Solutions with $T'=0$}

So far we found only solution having $T=0$ everywhere. In this section we show that a solution with constant $T$ can be built for some specific $f(T)$ models. Again we restrict the analysis to the $R(r)=r$ case. Let us consider
\begin{align}
  a(r)=\ln(k\,r)-T_0\,r^2\,, \qquad b(r)=\ln 4 \,,
  \label{013}
\end{align}
with $k$ and $T_0$ constants, corresponding to the metric
\begin{align}
  ds^2= k\,r\,e^{-T_0r^2} dt^2 - 4 dr^2 - r^2d\Omega^2\,.
\label{014}
\end{align}
Again we notice that the factor in front of $dr^2$ is not of physical importance. The solution~(\ref{013}) immediately implies $T=T_0$. The field equations~(\ref{field:toff})-(\ref{field:thetaoff}) become
\begin{align}
  4\pi\rho &= -\frac{1}{4}f(T_0)+\left(\frac{T_0}{4}-\frac{3}{8r^2}\right)f_T(T_0) \,,\\
  4\pi p &= \frac{1}{4}f(T_0)+\frac{1}{4r^2}f_T(T_0) \,,\\
  4\pi p &= \frac{1}{4}f(T_0)-\frac{(1-2T_0\,r^2)^2}{32r^2}f_T(T_0)\,.
\end{align}
In order to satisfy the last two of these equations, we must require $f_T(T_0)=0$. This is a constraint over all the possible $f(T)$ models which implies that only some specific models among these allow the solution~(\ref{014}). From the field equations we obtain constant energy density and pressure,
\begin{align}
  p=-\rho=\frac{f(T_0)}{4} \,.
\end{align}

As an example of a $f(T)$ model which allows for this solution we consider probably the simplest one:
\begin{align}
  f(T)=T-\frac{T^2}{2T_0} \,.
\end{align}
One can easily verify that $f_T(T_0)=0$. For this model the energy density and pressure become $p=-\rho=T_0/8$. Note that requiring a positive energy density constrains $T_0<0$, while in the approximation $T\ll T_0$ standard teleparallel gravity (TEGR) is recovered.

\subsection{Polynomial Solutions}

Finally, we present a class of solutions which involves a non constant $T$ and works for every model of the form $f(T)=T^N/N$. As before, we perform the analysis with $R(r)=r$.

Consider the metric
\begin{align}
  ds^2= (k\,r)^mdt^2-\frac{1}{n^2}dr^2-r^2d\Omega^2\,,
  \label{016}
\end{align}
with $n$ and $m$ (real) numbers and $k$ a constant of dimension 1/length. The factor $1/n^2$ in front of $dr^2$ is physically unimportant. Metric~(\ref{016}) corresponds to choosing the two functions
\begin{align}
  a(r)=m\ln(k\,r)\,, \qquad\mbox{and}\qquad b(r)=2\ln n \,,
\end{align}
which yield the non constant tensor scalar
\begin{align}
  T=\frac{2 (n-1) (-m+n-1)}{n^2 r^2} \,.
  \label{015}
\end{align}
Of course, we have to constrain $n\neq 0$, $n\neq 1$ and $n\neq m+1$ in order to obtain a non-zero and regular $T$. Subtracting field equations~(\ref{011}) and~(\ref{012}) leads to
\begin{align}
  \frac{(n-1) (m-2n+2) (m-n+1)}{n^4 r^4} f_{TT} + 
  \frac{\left(m^2-4 m+4 n^2-4\right) f_T}{8 n^2 r^2}=0 \,.
\end{align}
At this point we use expression~(\ref{015}) in order to replace $r^2$ with $T$ in this last equation. In this manner we obtain a relatively simple differential equation for the function $f(T)$
\begin{align}
  \left(m^2-4 m+4 n^2-4\right)f_T-4(m-2n+2)\,T\,f_{TT}=0 \,.
\end{align}
If $m\neq 2n+2$ the solution is
\begin{align}
  f(T)=\gamma+\frac{\xi}{N}T^N \,, \qquad\mbox{with}\qquad N = \frac{m^2+4(n-1)^2}{4(2+m-2n)} \,,
  \label{017}
\end{align}
and $\gamma$, $\xi$ constants.
Thus, for all these $f(T)$ models metric~(\ref{016}) represents a spherically symmetric solution. Note that TEGR is recovered when $N=1$, which can happen for a wide choice of the parameters $n$ and $m$. For example, setting $n=\sqrt 2$ and $m=2$ gives $N=1$.

Finally, from the field equations we can read off the expressions for the energy density and pressure
\begin{align}
  16 \pi\rho &= -\gamma+\xi\,h_1(n,m) \left(\frac{2(n-1) (-m+n-1)}{n^2 r^2}\right)^N \,,\\
  16 \pi p &=\gamma+\xi\, h_2(n,m) \left(\frac{2(n-1) (-m+n-1)}{n^2 r^2}\right)^N \,,
\end{align}
where $h_1$ and $h_2$ are two constants, depending on the parameters $n$ and $m$ in a complicated manner. If $N>0$ when $r\rightarrow\infty$ we have $p=-\rho=\gamma$, while if $N<0$ this happens at $r=0$. The explicit forms of $\rho$ and $p$ do depend upon the choice of $n$ and $m$. In general, we can state that a necessary condition to have both pressure and energy density positive everywhere is to set $\gamma=0$.

Thus, we were able to build a class of solutions presenting a non-constant $T$, allowed by a large number of $f(T)$ models of the form~(\ref{017}). These models are of general interest in cosmology and astrophysics since their study could lead to a better understanding of specific problems.

\section{Conclusions}

The objective of this paper was to show that relativistic stars exist in $f(T)$ modified gravity and to derive some static and spherically symmetric perfect fluid solutions explicitly. Our starting point was a careful study of the field equations and the resulting conservation equations for two choices of the tetrad. We showed that the conservation equation of $f(T)$ conceptually differs from those of General Relativity and its teleparallel equivalent. In doing so we also suggested a natural gauge current for $f(T)$ modified gravity. We derive the energy-momentum conservation equation by using two different approaches and both these results agree.

To begin with the study of exact solutions, we firstly studied some very simple models using a diagonal tetrad, by considering a constant torsion scalar $T'=0$. Next, we examined some well known solutions of General Relativity, including Schwarzschild type solutions and the Einstein static universe. It turns out that the $f(T)$ field equations for a Schwarzschild type ansatz of the metric are very difficult to solve and we were not able to make progress in this direction. We also found that the Einstein static universe with the usual assumptions of constant energy density and pressure only exists for the trivial $f(T)=T$, which made it necessary to look for generalisations. A diagonal tetrad, though appealing, poses stringent constraints on possible solutions. 

We thus considered a non diagonal tetrad adapted to spherically symmetry. For this choice of the tetrad, the conservation equation turns out to be equivalent to that of general relativity. We were able to construct a variety of exact solutions by making simplifying assumptions on the free functions. Among those solutions, we found one type of solution with a close resemblance to the Schwarzschild solution, we also found solutions with negative pressure. Finally, we derived new classes of solutions which we call polynomial solutions. This particular class of solutions has some very interesting properties. 

\section*{Acknowledgements}
We would like to thank Franco Fiorini for the useful discussions and comments on the manuscript. We would also like to thank Francisco Lobo for comments. AM is grateful to the organisers of PASCOS 2011 where parts of this work were presented.

\end{document}